\documentclass{raa_onecolumn}
\usepackage{graphicx,times}
\usepackage{natbib}
\usepackage{amssymb,amsmath}
\bibpunct{(}{)}{;}{a}{}{,}
\usepackage{epstopdf}

\begin{document}

   \title{Overview of the LAMOST-{\sl Kepler} project}

 \volnopage{ {\bf 2020} Vol.,\ {\bf 20} No. {\bf XX}, 000--000}
   \setcounter{page}{1}

\author{J.N. Fu\inst{1}\footnote{Corresponding authors}, P. De Cat\inst{2}, W. Zong\inst{1}{$^{\star\star}$}, A. Frasca\inst{3}, R.O. Gray\inst{4}, A.B. Ren\inst{5}, J.~Molenda-{\.Z}akowicz\inst{6}, C.J. Corbally\inst{7}, G. Catanzaro\inst{3}, J.R. Shi\inst{8}, A.L. Luo\inst{8}, and H.T. Zhang\inst{8}}

   \institute{Department of Astronomy, Beijing Normal University, Beijing~100875, P.~R.~China;
   {\it jnfu@bnu.edu.cn; weikai.zong@bnu.edu.cn}\\
\and
Royal Observatory of Belgium, Ringlaan 3, B-1180 Brussel, Belgium\\
\and
INAF -- Osservatorio Astrofisico di Catania, Via S. Sofia 78, I-95123 Catania, Italy\\
\and
Department of Physics and Astronomy, Appalachian State University, Boone, NC 28608, USA
\and
Department of Astronomy, China West Normal University, Nanchong 637002, China
\and
Astronomical Institute of the University of Wroc\l{}aw, ul. Kopernika 11, 51-622 Wroc\l{}aw, Poland
\and
Vatican Observatory Research Group, University of Arizona, Tucson, AZ 85721-0065, USA
\and
Key Lab for Optical Astronomy, National Astronomical Observatories, Chinese Academy of Sciences, Beijing 100012, P.~R.~China\\
\vs \no
   {\small Received 2020 August 15; accepted 2020 XXX}
}

\abstract{The NASA {\sl Kepler} mission obtained long-term high-quality photometric observations for a large number of stars in its original field of view from 2009 to 2013.
In order to provide reliable stellar parameters in a homogeneous way, the LAMOST telescope began to carry out low-resolution spectroscopic observations for as many stars as possible in the {\sl Kepler} field in 2012.
By September 2018, 238,386 low-resolution spectra with SNR$_g \geq 6$ had been collected for 155,623 stars in the {\sl Kepler} field, enabling the determination of atmospheric parameters and radial velocities, as well as spectral classification of the target stars.
This information has been used by astronomers to carry out research in various fields, including stellar pulsations and asteroseismology, exoplanets, stellar magnetic activity and flares, peculiar stars and the Milky Way, binary stars, etc. We summarize the research progress in these fields where the usage of data from the LAMOST-{\sl Kepler} (LK) project has played a role. In addition, time-domain medium-resolution spectroscopic observations have been carried out for about 12,000 stars in four central plates of the {\sl Kepler} field since 2018.
The currently available results show that the LAMOST-{\sl Kepler} medium resolution (LK-MRS) observations provide qualified data suitable for research in additional science projects including binaries, high-amplitude pulsating stars, etc.
As LAMOST is continuing to collect both low- and medium-resolution spectra of stars in the {\sl Kepler} field, we expect more data to be released continuously and new scientific results to appear based on the LK project data.
}

   \authorrunning{{\it Fu et al.}: Overview of the LK project}            
   \titlerunning{{\it Fu et al.}: Overview of the LK projects}  
   \maketitle

%
\section{Motivations and Aims}           
\label{sect:intro}
The NASA space mission {\sl Kepler} was designed to detect Earth-like planets around solar-type stars with the transit method by monitoring continuously the brightness of around 200,000 stars in a fixed Field-of-View (FOV) of 105 square degrees in the constellations of Lyra and Cygnus \citep{Borucki2010}. It collected almost continuous time-series ultra-high precision photometry for the target stars in this FOV between May 2 of 2009 and May 11 of 2013, providing unprecedented light curves for not only the detection of exoplanets but also research in more extensive areas, including stellar oscillations, eclipsing binaries, stellar activity, star clusters, etc \citep{Barentsen2018}. However, in order to fully exploit the excellent data of {\sl Kepler} for many kinds of scientific research, knowledge of the basic physical parameters of the target stars is essential. These physical parameters include the effective temperature ($T_{\rm eff}$), surface gravity ($\log g$), metallicity ([M/H]), and projected rotation velocity ($v\sin i$), the latter being especially important for asteroseismic studies of oscillating stars \citep{Michel2006,Cunha2007}. Unfortunately, the {\sl Kepler} wide-passband photometry is not suitable for deriving stellar atmospheric parameters. Hence, ground-based photometric observations were carried out before the launch of {\sl Kepler} and led to the determination of stellar parameters recorded in the {\sl Kepler} Input Catalog \citep[KIC;][]{Brown2011}.  The precision of those parameters has proven to be too low in general for asteroseismic modelling \citep{Molenda2010, McNamara2012}. In addition, reliable information about stellar metallicity and rotation rate was lacking.
In view of the large number of target stars, a facility capable of making spectroscopic observations for multiple targets efficiently was very desirable.

The Large Sky Area Multi-Object Fiber Spectroscopic Telescope ($\rm{LAMOST}$, also called the Guo Shou Jing Telescope) has proven to be an ideal instrument for such aims, since it combines a large aperture of 3.6-4.9~m and a wide circular FOV with a diameter of $5\deg$ \citep{Wang1996}.The focal plane can host up to 4000 fibers, which are connected to 16 identical spectrographs \citep{Xing1998}. Since 2012 the LAMOST spectrographs have employed entrance slits with widths two-thirds of the diameter of the fibers (corresponding to $\sim 213\mu$m, or $2.2"$ on the sky) .  The spectral resolutions with those slits are $\sim1800$ and $\sim7500$ for the low- and medium-resolution gratings, respectively \citep[see more details in][]{Cui2012,Hou2018}.

In 2010, we launched the LK project. LAMOST observations to collect low-resolution spectra for as many stars as possible in the {\sl Kepler} FOV were started in May 2011 as phase~I of the project.  However, changes in the configuration of the instrument
were made in early 2012, and so none of the spectra obtained in 2011 are used in
our analysis.
Time-series medium-resolution spectroscopic observations, beginning in September 2018, are being obtained for stars in four central LAMOST plates in the {\it Kepler} field as phase~II.
The low-resolution spectra obtained in phase~I allow not only homogeneous determination of the stellar atmospheric parameters ($T_{\rm eff}$, $\log g$ and [Fe/H]) and spectral classification of the observed stars, but also estimation of the radial velocity ($RV$) and, in the case of rapid rotation, the projected rotational velocity ($v \sin i$). The higher resolution spectra enable the derivation of chemical abundances and the measurement of the strength of the H$\alpha$ chromospheric emission for stars with a lower level of magnetic activity.
For studying binaries and in particular large-amplitude variable stars, it is helpful to have more precise stellar parameters and time domain measurements of radial velocities. 
These were provided in phase~II by medium resolution spectra of the targets at multiple epochs. Since the first data release of the LK project in 2015, the data obtained have been used for research in various fields with fruitful scientific results.

In this paper, we focus on phase~I of the LK project in sections~2-5. The survey design of the LK project of phase~I is described in section~\ref{sect:2}. We summarize the observations and spectra in section~\ref{sect:3}. Section~\ref{sect:4} is devoted to the description of three computer codes applied to estimate the stellar parameters, radial velocity, projected rotational velocity, and spectral classification. Scientific research based on the data of phase~I of the LK project is presented in section~\ref{sect:5}. In sections~\ref{sect:6} and \ref{sect:conclusions}, we discuss phase~II of the LK project and give a brief summary and prospects.

\section{Survey Design}           
\label{sect:2}

Since the {\sl Kepler} FOV is approximately a quadrilateral with an area of 105 square degrees, while LAMOST has a circular FOV of 20 square degrees, we used 14 LAMOST plates to almost fully cover the {\sl Kepler} FOV; these are called "LK-fields". Each of the LK-fields has a star brighter than $V$ = 8 at the center, as designated by their plate ID (Figure~\ref{Fig2}). One can see the details of the 14 LK-fields in Figure~2 and Table~1 of \citet{De Cat2015}.

With the information for the stars in the {\sl Kepler} FOV provided by KIC, we divided the stars into ``Standard targets'', ``$\rm{KASC}$ targets'', ``Planet targets'' and ``Extra targets'' groups which correspond, respectively, to $\sim$120 MK secondary standard stars, $\sim$6500 targets selected by the {\sl Kepler} Asteroseismic Science Consortium (KASC), $\sim$150,000 targets selected by the {\sl Kepler} planet search group \citep{Batalha2010}, and $\sim$1,000,000 targets from the KIC \citep{Brown2011}. Observing priorities were assigned to those four groups, in that order, from high to low.

With the target lists of the 14 LK-fields, the code ``Survey Strategy System'' was used to prepare the observation plans of the LK project to optimize the effective use of the fibers \citep{Cui2012}.


\begin{table}
\centering
\caption[]{General contents of the LK project observations during the regular survey phase from 2012 September to 2018 June. \label{tab1}}
\setlength{\tabcolsep}{15pt}
\fns
 \begin{tabular}{cccrr}
  \hline\noalign{\smallskip}
Year& LK field & Plate & Spectra & Parameter \\
  \hline\noalign{\smallskip}
2012 & 3  &   7  & 17\,659  & 11\,682    \\
2013 & 6  &  14  & 39\,309  & 28\,115    \\
2014 & 7  &  14  & 38\,516  & 29\,351    \\
2015 & 11 &  32  & 97\,247  & 81\,381    \\
2017 & 7  &  18  & 40\,763  & 28\,232    \\
2018 & 1  &  2   & 4\,892   & 3\,957    \\
\hline
Total  &   &     & 238\,386  & 182\,618 \\
Unique &   &     & 100\,219  &  85\,932 \\
2$\times$ &  &   &  37\,563  &  28\,555 \\
3$\times$ &  &   &  12\,343  &   8\,205 \\
4$\times$ &  &   &   3\,441  &   2\,321 \\
+5$\times$ & &   &   2\,057  &   1\,016 \\
\noalign{\smallskip}\hline
\end{tabular}
~\\
~\\
\footnotesize{The number of multiple revisited targets depends on the criteria one chooses when performing cross-identification.}
\end{table}

\section{Observations and Spectral Reductions}
\label{sect:3}
Observations on the LK project began on May 30, 2011 during the pilot survey phase of LAMOST. As the data quality improved with the adoption of the narrower slit width in June 2012, only the observations obtained since that date have been retained for the LK project.  Table~\ref{tab1} lists the LK project observations from June 2012 to June 2019 June.

\begin{figure}
   \centering
   \includegraphics[width=10.0cm, angle=0]{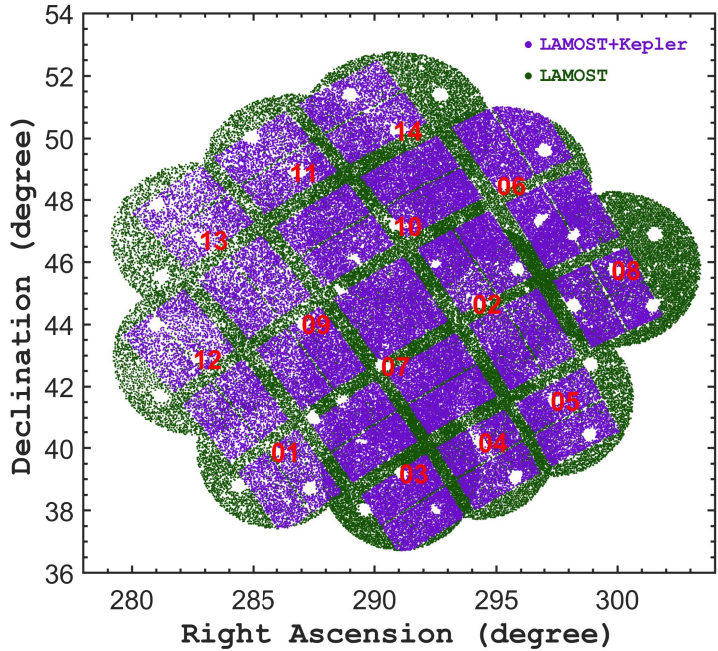}
   \caption{Sky coverage of all targets observed during phase\,I of the LK project. The stars observed by LAMOST and with {\sl Kepler} photometry are purple, while others are dark green. The numbers in red mark the central positions of the 14 LK fields. }
   \label{Fig2}
\end{figure}

\begin{figure}
   \centering
   \includegraphics[width=10.0cm, angle=0]{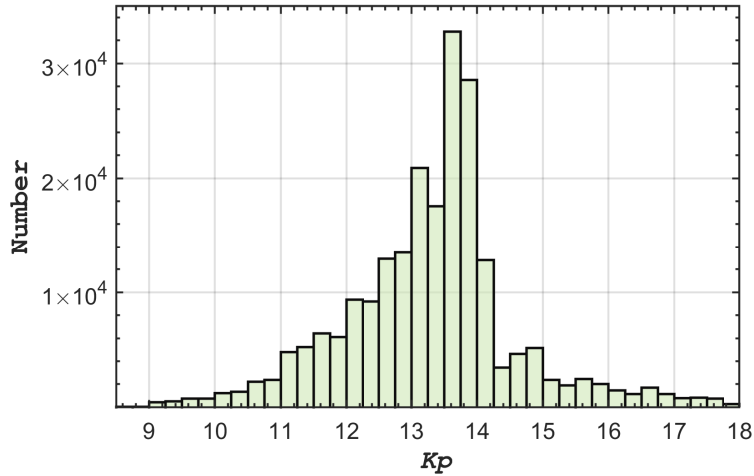}
   \caption{Distribution of the {\sl Kepler} magnitude of the stars observed by LAMOST during phase\,I of the LK project.}
   \label{Fig3}
\end{figure}

\begin{figure}
   \centering
   \includegraphics[width=10.0cm, angle=0]{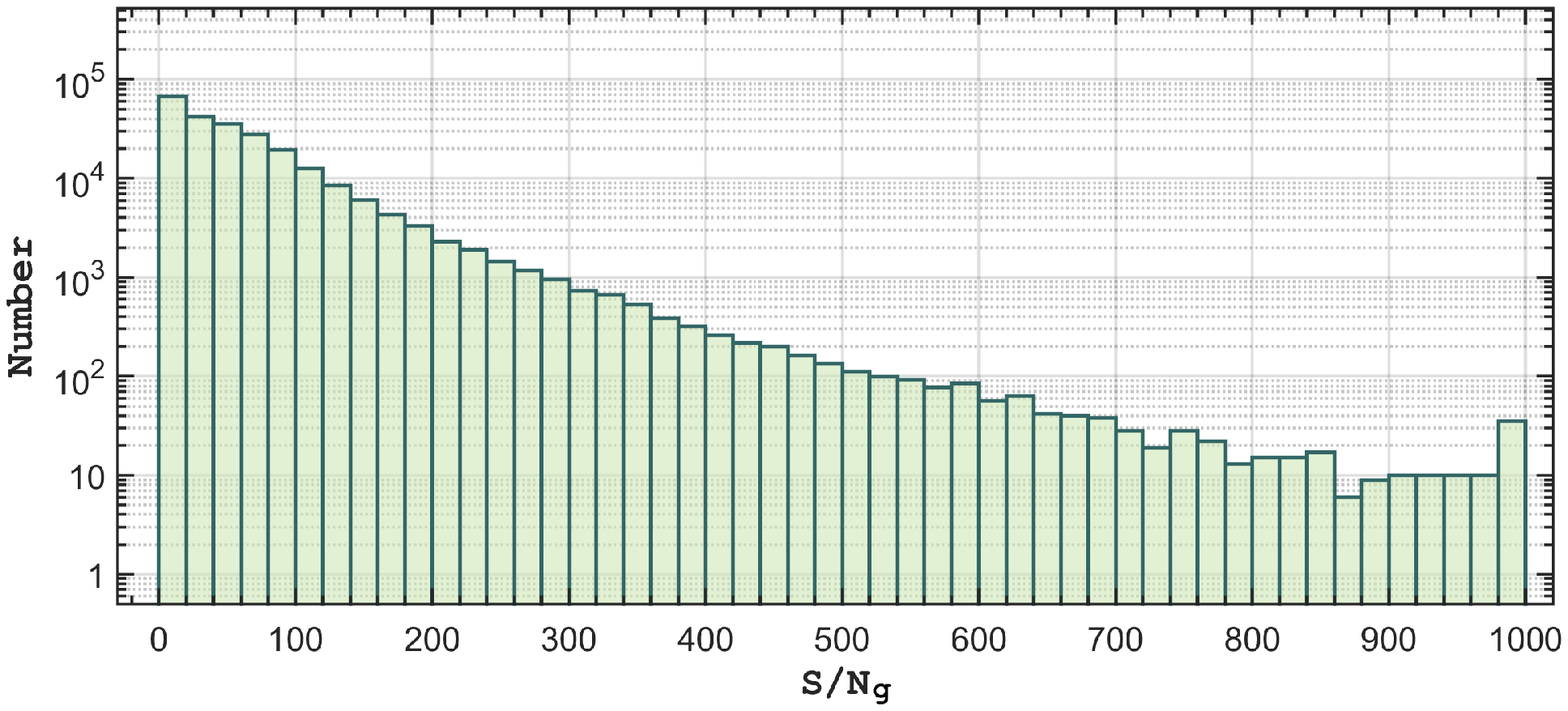}
   \caption{Distribution of the SNR in the $g$ band for the low-resolution spectra of the LK project.}
   \label{Fig4}
\end{figure}

The raw spectroscopic data were reduced with the LAMOST 2D and 1D pipelines \citep{Luo2012,Luo2015}. Those pipelines together provide wavelength and flux-calibrated spectra. Those spectra with Signal-to-Noise ratios (SNR) in the {\sl SDSS g} band higher than or equal to 6 are accepted as "qualified" data. By the time of the LAMOST Data Release~7 (DR7; March 2020), we had obtained in total 238,386 qualified spectra of 155,623 stars. Figure~\ref{Fig2} shows the sky coverage of all targets observed by the LK project. Figures~\ref{Fig3} and \ref{Fig4} show the distribution of {\sl Kepler} magnitudes of the stars observed by LAMOST during phase\,I of the LK project and the distribution of the $g$-band SNR of the low-resolution spectra of the LK project, respectively.

\section{Stellar Parameter Determination}
\label{sect:4}
For all the spectra obtained for the LK project, three teams have been using independent approaches to characterize the observed target stars and derive stellar parameters.

\subsection{LASP}
\begin{figure}
   \centering
   \includegraphics[width=7.0cm, angle=0]{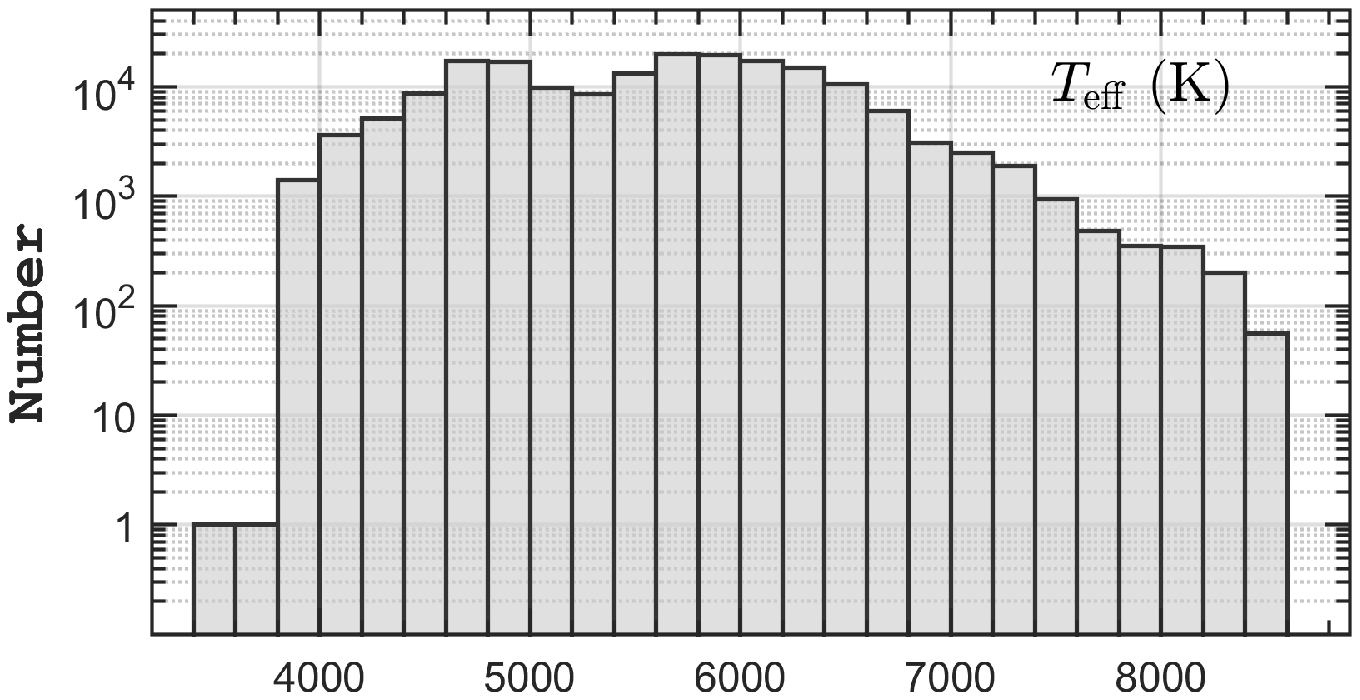}
   \includegraphics[width=7.0cm, angle=0]{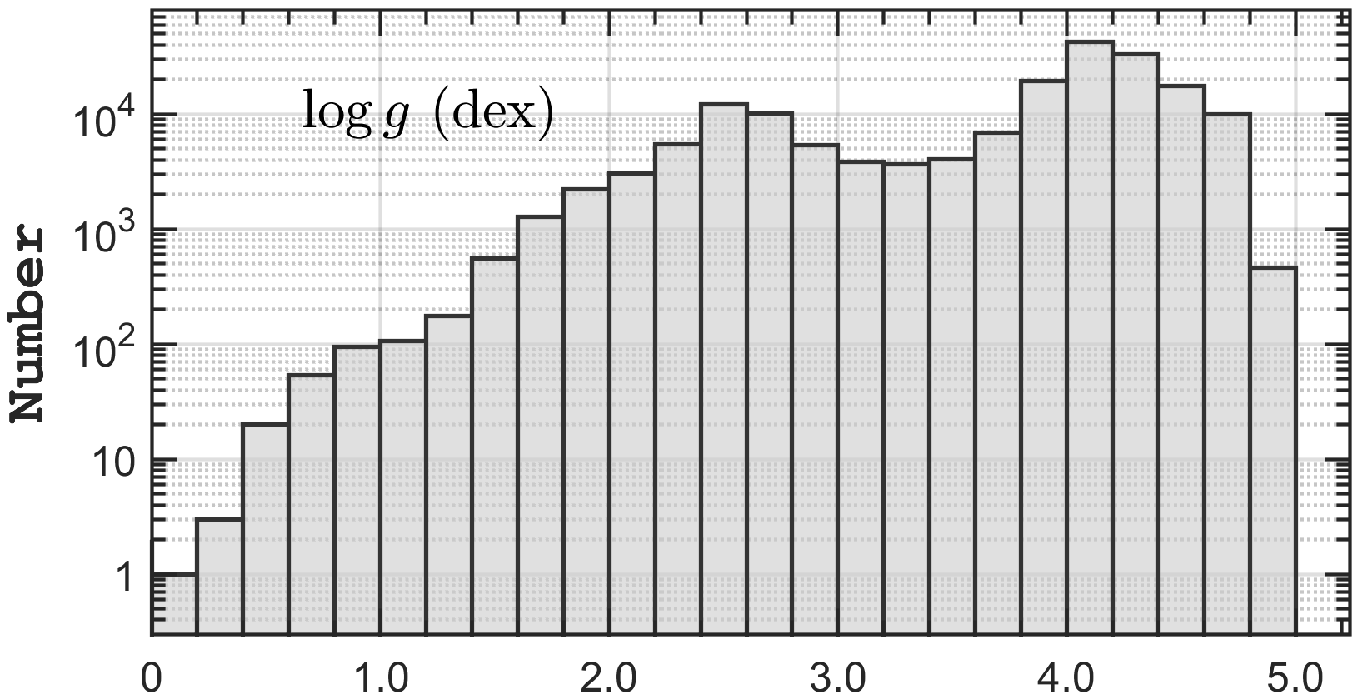}\\
   \includegraphics[width=7.0cm, angle=0]{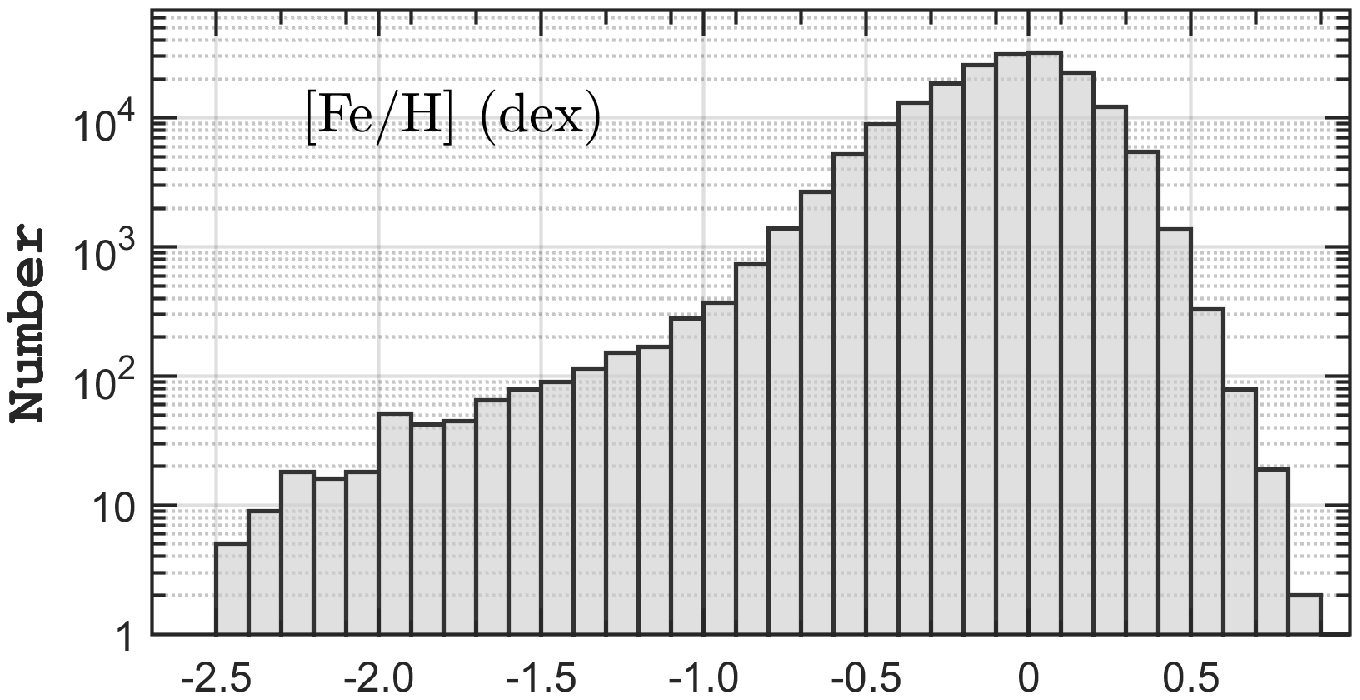}
   \includegraphics[width=7.0cm, angle=0]{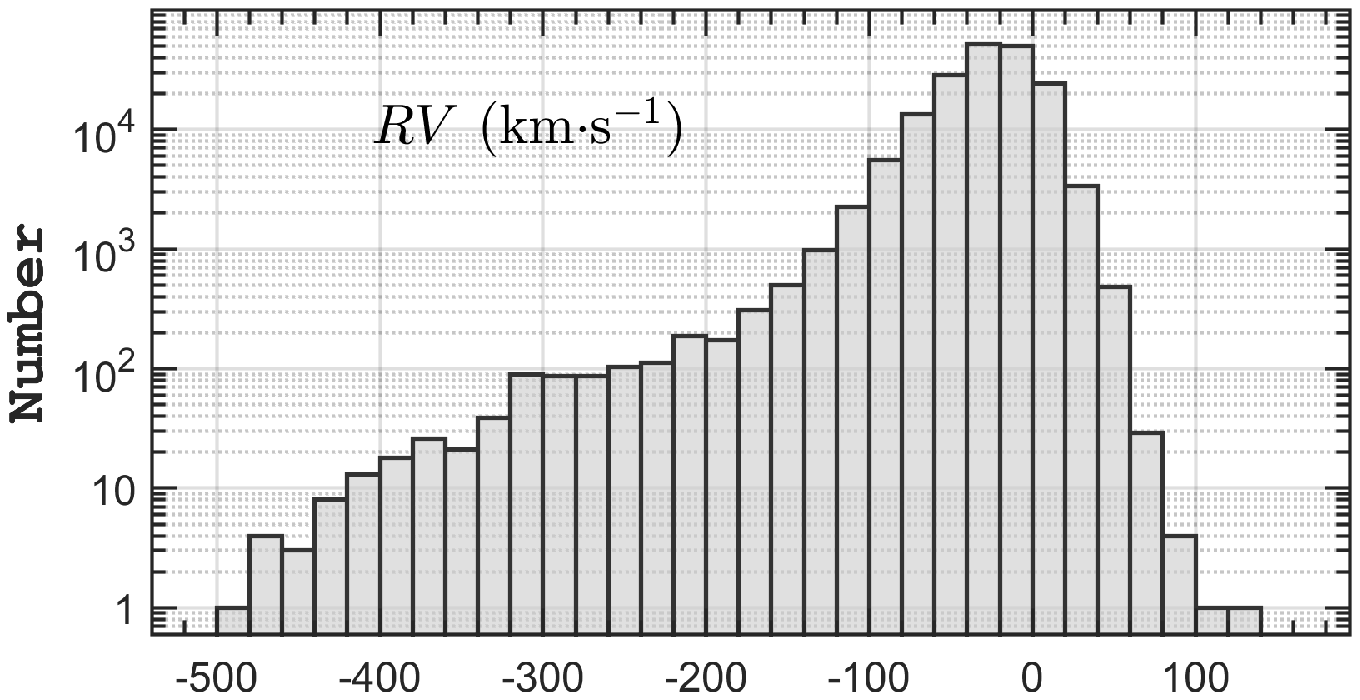}
   \caption{Histogram of atmospheric parameters and $RV$ derived from 238,386 spectra. Top left panel: the effective temperature $T_{\rm eff}$ (K); top right: the surface gravity $\log g$ (dex); bottom left: the metallicity [Fe/H] (dex); and bottom right: the radial velocity $RV$ (km s$^{-1}$).}
   \label{Fig5}
   \end{figure}

\begin{figure}
   \centering
   \includegraphics[width=10.0cm, angle=0]{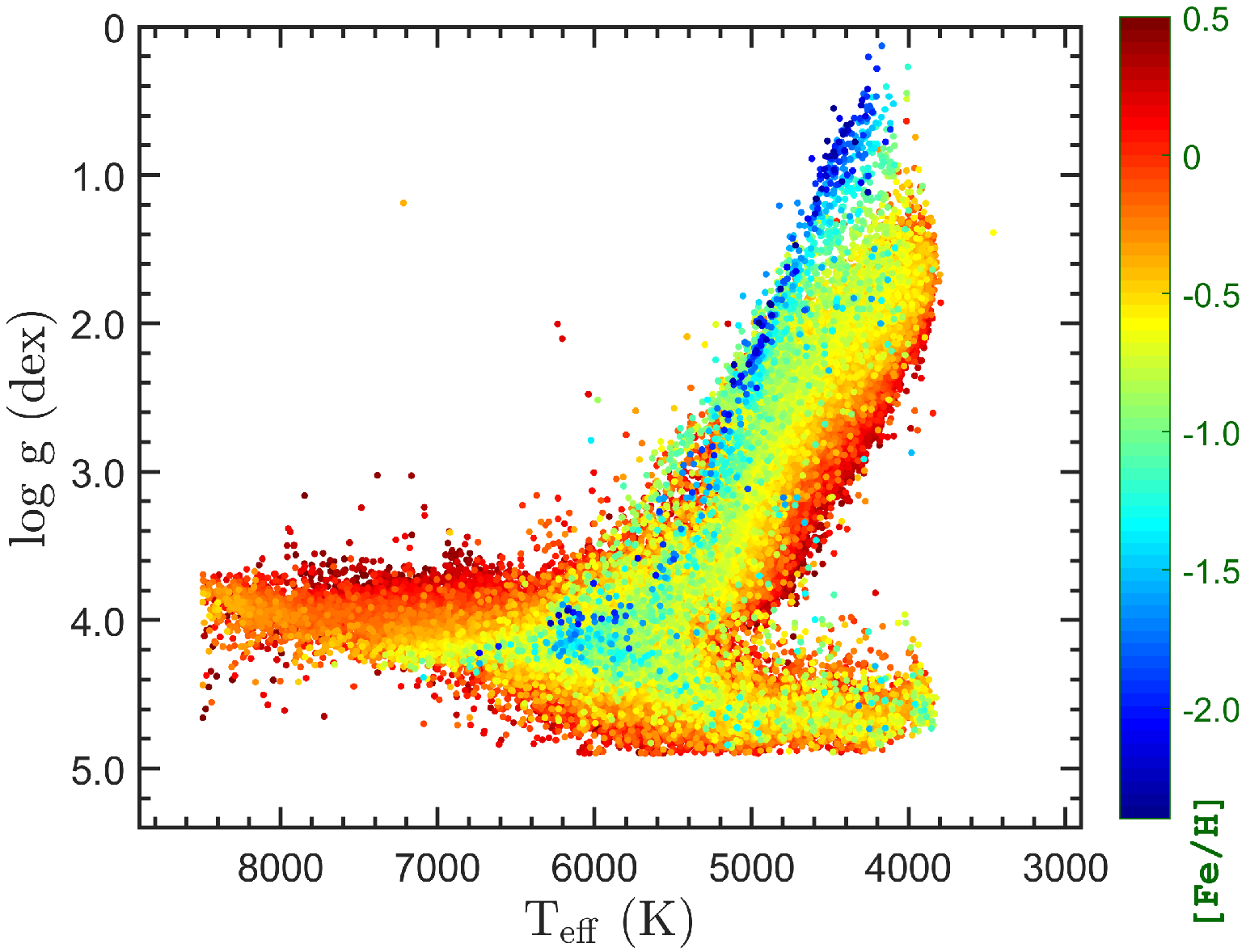}
   \caption{Kiel diagram ($\log g$ vs. $T_{\rm eff}$) of the "qualified" LK spectra. The parameters are derived from the LASP pipeline. Note that different colors indicate different values of metallicity [Fe/H].}
   \label{Fig6}
   \end{figure}

The ``Asian team'' performed statistical analyses of the stellar parameters resulting from the LAMOST stellar parameter pipeline \citep[LASP;][]{Luo2015}. The LASP code determines the stellar atmospheric parameters ($T_{\rm eff}$ , $\log g$, [Fe/H]) and $RV$ for late A-, F-, G-, and K-type dwarf and giant stars. \citet{Ren2016a} and \citet{Zong2018} analyzed the stellar parameters and $RV$ of the LK project provided by LASP in DR3 and DR5 of LAMOST, respectively. DR7 of LAMOST contains the latest stellar parameter determinations and $RV$ for all of the LK project spectra from LASP. Figures~\ref{Fig5} and \ref{Fig6} show the histograms of these measurements and the Kiel diagram ($\log g$ vs. $T_{\rm eff}$) of the qualified LK spectra, respectively.

\subsection{ROTFIT}

The ``European team'' determined stellar parameters and spectral classification with an adapted version of the code ROTFIT \citep{Frasca2003,Frasca2006}. \citet{Frasca2016} derived the $RV$ and atmospheric parameters for 61,753 spectra of 51,385 target stars of the LK project, identified interesting and peculiar objects, such as stars with $RV$ variations, ultrafast rotators, and emission-line objects. In addition, 442 chromospherically active stars were discovered in the {\sl Kepler} field.

\subsection{MKCLASS}

The ``American team'' developed the code MKCLASS for automatically classifying stars with the spectra obtained by the LK project on the MK spectral classification system independent of the stellar parameter determination \citep{Gray2014}. \citet{Gray2016} presented the quality and reliability of the spectral types of 80,447 stars with 101,086 spectra, computed the proportion of A-type stars that are Am stars, and identified 32 new barium dwarf candidates in the {\sl Kepler} field.

\section{Scientific Research}
\label{sect:5}

\begin{figure}
   \centering
   \includegraphics[width=10cm, angle=0]{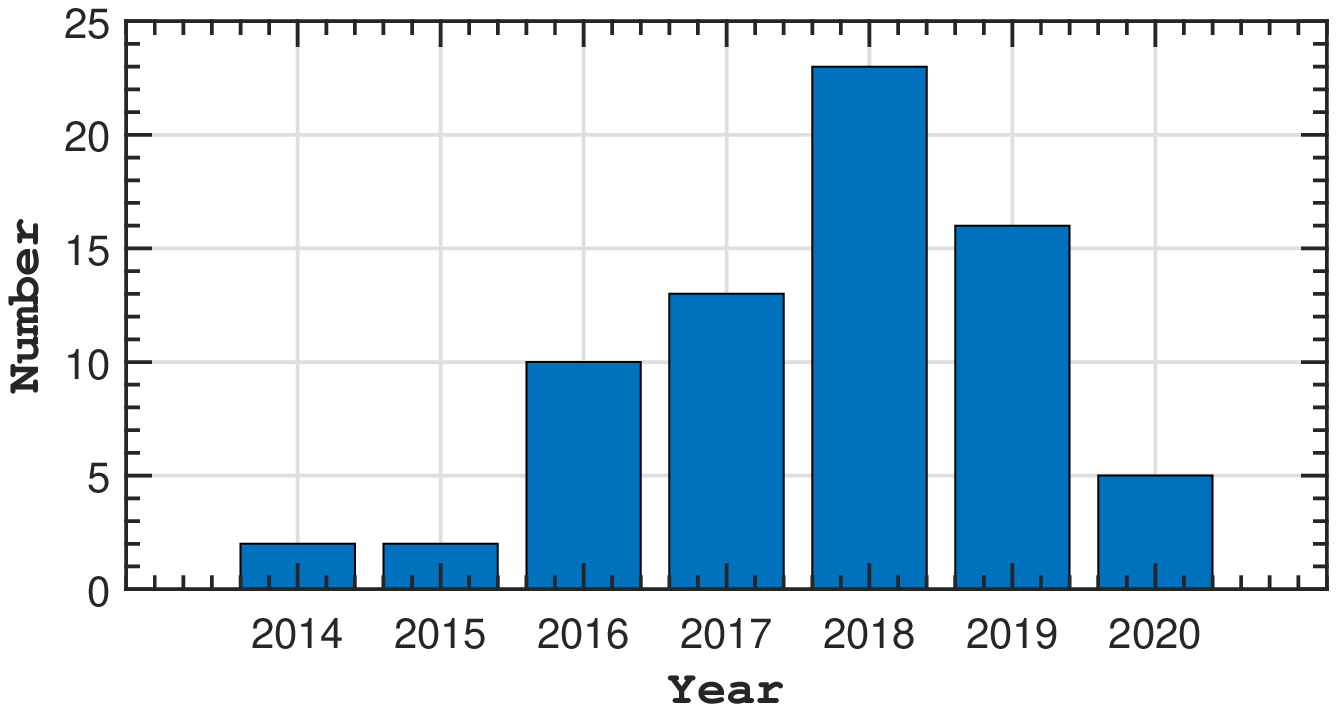}
   \caption{Number of refereed publications using LK project data according to year.}
   \label{Fig7}
   \end{figure}

Spectra and stellar parameters derived by the LK project have been used by astronomers in various research fields. In the literature, we have to date found 70 refereed publications, which is certainly a lower limit, and a similar number of non-refereed articles that use data from the LK project. Here, we summarize only the research  published in refereed papers and based on the low-resolution spectra observed by LAMOST for the stars in the {\sl Kepler} field. Figure~\ref{Fig7} shows the number of refereed publications in different years from 2014 to 2020.

\begin{figure}
   \centering
   \includegraphics[width=10.0cm, angle=0]{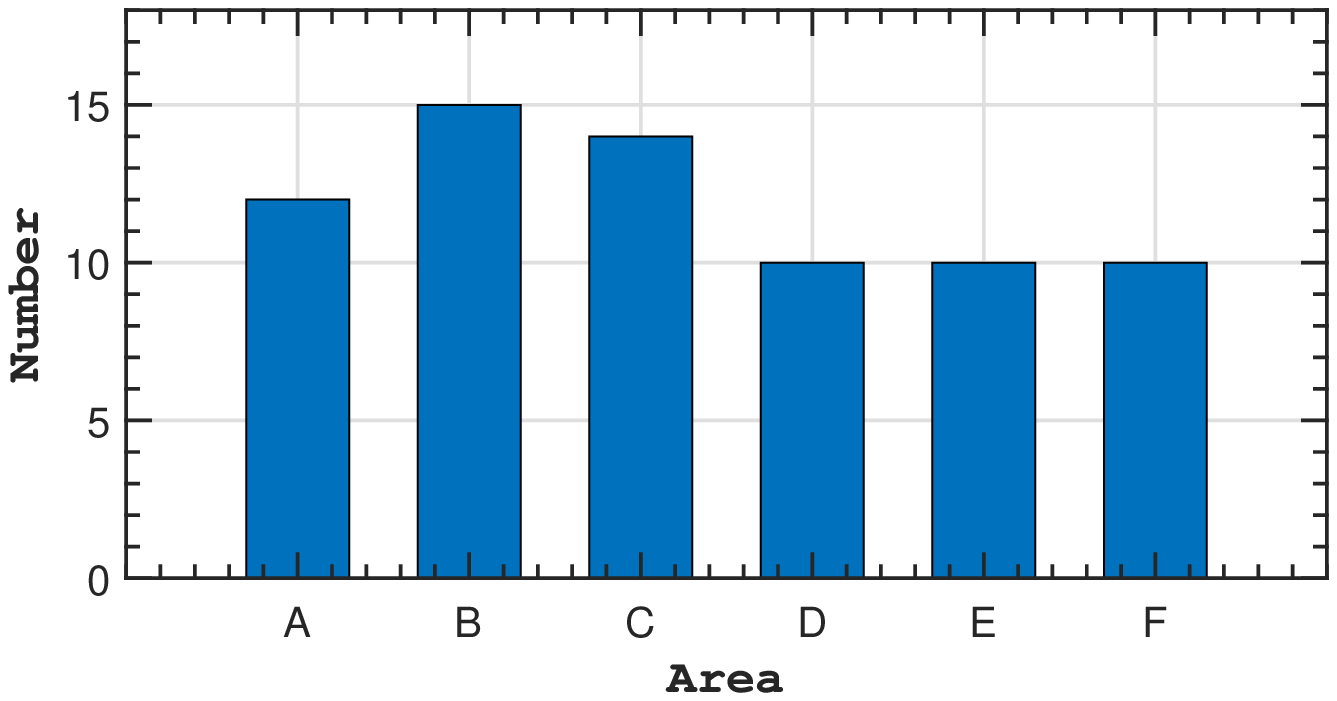}
   \caption{Number of refereed publications using LK project data according to area. The letters denote the following categories:  A: survey spectra and stellar parameter determination; B: stellar pulsations and asteroseismology; C: exoplanets; D: stellar magnetic activity and flares; E: peculiar stars and the Milky Way; F: binary stars.}
   \label{Fig8}
   \end{figure}

We classify the refereed papers into six research areas: A) survey spectra and stellar parameter determination; B) stellar pulsations and asteroseismology; C) exoplanets; D) stellar magnetic activity and flares; E) peculiar stars and the Milky Way, and F) binary stars. We show the number of refereed papers according to the research area in Figure~\ref{Fig8}.

\subsection{Survey Spectra and Stellar Parameters}
Up to the time of writing, there have been five refereed publications that have (i)  introduced the LK project, (ii) released the survey data, (iii) carried out spectral classifications and, (iv) derived and calibrated stellar parameters for up to 156,390 stars in the {\sl Kepler} field \citep[][respectively]{De Cat2015,Frasca2016,Gray2016,Ren2016a, Zong2018}.
Using 12,000 stars observed by the LK project, \citet{Dong2014} showed that the metallicities of {\sl Kepler} field stars reported in the KIC were systematically underestimated and spanned a range of values smaller than that displayed by the [Fe/H] values derived spectroscopically by the LK survey.
On the other hand, \citet{Liu2015} used the asteroseismology-determined surface gravities of the giant stars based on the {\sl Kepler} light curves to calibrate the measurements derived by the LAMOST pipeline. \citet{Wang2016} applied the $T_{\rm eff}$ and [Fe/H] values of the LK project to the global oscillation parameters to establish empirical calibration relations for the $\log g$ values of dwarfs and giants. \citet{Xiang2017} estimated stellar atmospheric parameters, absolute magnitudes, and elemental abundances from the LAMOST spectra with the LAMOST Stellar Parameter Pipeline at Peking University (LSP3). In addition, the stellar properties of {\sl Kepler} targets derived from LK project data were discussed by \citet{Mathur2017}, \citet{Pande2018}, and \citet{Scaringi2018}.

\subsection{Stellar Pulsations and Asteroseismology}
The LK project data have been applied to studies of stellar pulsations. \citet{Hey2019} discovered six roAp stars in the {\sl Kepler} long-cadence data with four of them identified as chemically peculiar stars using LAMOST spectra. \citet{Smalley2017} studied a large sample of A and Am stars with spectral types from LAMOST and light curves of WASP, finding that the amplitude distributions agree with results obtained from {\sl Kepler} photometry. \citet{Bhardwaj2019} studied Mira variables in the Magellanic Clouds compared with those in the {\sl Kepler} field, and used the spectra provided by the LK project. \citet{Balona2018} illustrated the effect of tides on self-driven stellar pulsations of the heartbeat $\delta$ Scuti variable KIC~4142768 using the LK project data.

\citet{Yu2016} analyzed solar-like oscillations in 1523 {\sl Kepler} red giants which had previously been misclassified as subgiants due to large errors in the KIC, by comparing the KIC surface gravities with the values derived by the LK project. \citet{Yu2018} characterized solar-like oscillations and granulation for 16,094 oscillating red giants by using the long-cadence data of {\sl Kepler}, which are helpful to lift degeneracies in deriving atmospheric parameters from LAMOST. \citet{Li2017} performed asteroseismic analysis on six solar-type stars observed by {\sl Kepler} with the LK project atmospheric parameters serving as constraints on stellar models. \citet{Li2018} conducted an analysis for the asteroseismic binary system KIC~7107778 with a non-eclipsing unresolved companion having solar-like oscillations. The LK project atmospheric parameters were used when the two stars were modelled theoretically.

LAMOST observations for stars in the {\sl Kepler} field help to derive more precise parameters for the target stars of asteroseismic analysis with the {\sl Kepler} time-series photometric data. \citet{Ren2016b} found that the surface gravities yielded by LSP3 are in good agreement with the asteroseismic values. \citet{Wu2017, Wu2018, Wu2019} examined the age determination as well as the contamination rate for a sample of 150 main sequence turn-off stars, estimated masses and ages of 6940 red giant branch stars, and presented a catalogue of stellar age and mass estimates for a sample of 640,986 red giant branch stars, respectively. Using spectra of the LK project to investigate  chromospheric activity and {\sl Kepler} light curves to investigate photospheric activities of 2603 stars, Zhang et al. (2020a) found that 254 stars with near-solar rotation periods had chromospheric activities systematically higher than those with undetected rotation periods. \citet{Peralta2018} reported a new method for extracting seismic indices and granulation parameters for more than 20,000 CoRoT and {\sl Kepler} red giants. \citet{Stassun2018} presented empirical accurate masses and radii of single stars with TESS and {\it Gaia} data, where the values of {\sl Kepler} single stars were compared.

\subsection{Exoplanets}
LK project data in the {\sl Kepler} field have been used successfully to investigate exoplanets and their parent stars.  For instance, \citet{Mulders2016} took a sample of over 20,000 {\sl Kepler} stars with spectroscopic metallicities from the LAMOST survey to explore how the exoplanet population depends on host star metallicity as a function of orbital period and planet size.

By using the precise spectroscopic parameters of host stars from LAMOST observations, \citet{Xie2016} measured the eccentricity distributions for a homogeneous sample of 698 planets discovered by {\sl Kepler}. They found that in systems with only one transiting planet, those planets are on eccentric orbits with e$\approx$0.3, whereas the planets in multi-planet systems have nearly circular and coplanar orbits similar to those of our solar system. \citet{Dong2018} used accurate stellar parameters for main-sequence stars provided by the LK project to study the distributions of short-period {\sl Kepler} planets as a function of host star metallicity, which helped them to discover a population of short-period, Neptune-size planets sharing key similarities with hot Jupiters.

\citet{Wang2018} presented a statistical study of the planet-metallicity correlation by studying 744 stars with candidate planets in the {\sl Kepler} field observed with LAMOST. The clues uncovered suggested that giant planets around FGK stars probably form through core accretion, with high metallicity as a prerequisite for massive planets to form. \citet{Petigura2017, Petigura2018} described the California-Kepler Survey which aims to improve our knowledge of the properties of stars found to host transiting planets by the {\sl Kepler} mission.  That survey used LK stellar parameters to define their observational sample, and by calibrating LK project metallicities to their own, effectively extended their sample.  They used those data to explore statistics of various types of planets as a function of metallicity.  \citet{Bashi2019} combined information from LAMOST and the California-Kepler Survey to explore the occurrence rate of small close-in planets among {\sl Kepler} target stars. Their results suggested there are two regions in the ([Fe/H], [$\alpha$/Fe]) plane in which stars tend to form and maintain small planets.

Regarding the {\sl Kepler} planet occurrence rates, \citet{Guo2017, Narang2018, Zhu2018}, \citet{Zhu2019}, and \citet{Hardegree2019} all reported research progress. \citet{Murphy2016} studied a planet orbiting an A-type star in the {\sl Kepler} field. \citet{Kawahara2019} provided a comprehensive catalog of transiting {\sl Kepler} planets near the snow line. In all the above publications, LK project data were applied in the analyses.

\subsection{Stellar Magnetic Activity and Flares}
The long-term almost-continuous photometric observations by {\sl Kepler} have provided an unprecedented opportunity for the study of magnetic activity at the photospheric level for a large number of stars. In particular, rotational modulation of the ultra-precise photometry enables detection of rotation periods, differential rotation, and starspot distributions.  Flares from a large number of stars may likewise be studied and characterized. {\sl Kepler} photometry along with information on the target stars derived from spectra obtained by facilities like LAMOST, has placed research in this field on a more solid foundation. For example, \citet{Yang2019} presented a flare catalog of the {\sl Kepler} mission, comprising 3420 flare stars and 162,262 flare events. The incidence of flare stars rises with decreasing temperatures, where the latter values were provided by the LK project.

With respect to solar-type stars, \citet{Karoff2016} analyzed observations made with LAMOST of 5,648 sources, including 48 superflare stars, finding that superflare stars are generally characterized by higher chromospheric activity levels than other stars including the Sun. 
There was significant evidence for superflares and solar flares to most likely share the same origin, and robust estimates of the relationship between chromospheric activity and the occurrence of superflares were presented. 
On the other hand, studies of flaring M dwarfs in the {\sl Kepler} field have been presented in several articles by using both satellite photometry and LAMOST data \citep{Yang2017, Chang2017, Chang2018, Lu2019}.
Using {\sl Kepler} and LK project data, research progress was reported on the chromospheric activity of periodic variable stars \citep{Zhang2018b} and 
long-rotation-period main-sequence stars \citep{Cui2019}. By extracting chromospheric indices for 59,816 stars from LAMOST spectra and photospheric index data for 5,575 stars from {\sl Kepler} light curves, \citet{Zhang2020b} studied the magnetic activity of F-, G-, and K-type stars in the {\sl Kepler} FOV.

\subsection{Peculiar Stars and The Milky Way}
A number of articles incorporating LK project data have been devoted to the Li-rich giants in the {\sl Kepler} field. \citet{Silva2014} used LAMOST observations in the {\sl Kepler} field to search for potential Li-rich candidates and found the first confirmed Li-rich core-helium-burning giant, as revealed by asteroseismic analysis. \citet{Kumar2018} reported two new super Li-rich K giants discovered on the basis of LK project data and subsequently confirmed with high-resolution spectra obtained with other observational facilities. The LK project spectroscopic data were used to refine the selection function recently derived by \citet{Casey2018} for giant stars in the {\sl Kepler} field. 
The photometric selection function was then employed to identify three metal-poor giant candidates whose masses, previously estimated from standard asteroseismic scaling relations, turned out to be overestimated by a factor of 20-175\,\%. \citet{Singh2019a} introduced a survey of Li-rich giants in the {\sl Kepler} field with LAMOST observations to determine their evolutionary phases. \citet{Singh2019b} reported the discovery of two new super Li-rich K giants in the {\sl Kepler} field in the red clump phase with core He burning, based on {\it Gaia} astrometry and secondary calibrations using {\sl Kepler} asteroseismic analyses and LAMOST spectroscopic data.

LK project data have been applied to the study of other types of stars and stellar populations..
\citet{Cook2017} developed a method to identify the spectroscopic signature of unresolved L-dwarf ultracool companions, employing LAMOST survey spectra and {\sl Kepler} light curves. \citet{Ho2017} measured carbon and nitrogen abundances for a large number of giant stars from their low-resolution LAMOST DR2 spectra, which were then used to infer stellar masses, and compared those masses with the masses of giants in the {\sl Kepler} field estimated from asteroseismology. In order to provide precise and accurate stellar parameters for Galactic archaeology, \citet{Mints2017} developed an unified tool UniDAM to estimate distances, ages, and masses from spectrophotometric data of different surveys including LAMOST, to obtain a homogenised set of stellar parameters which were verified with those derived from asteroseismic data of {\sl Kepler}.

\citet{Bell2017} assessed the photometric variability of nine stars with spectroscopic parameters from the extremely low-mass white dwarf survey using {\sl Kepler} photometry and LAMOST spectroscopic observations. With the help of LAMOST spectra, \citet{Zhang2018a} found four misclassified main-sequence B stars in the {\sl Kepler} field, and presented spectroscopic and frequency analyses of those four stars based on LAMOST spectra and {\sl Kepler} photometry.

\subsection{Binary Stars}
LK project data have also been used in the research of binary stars in the {\sl Kepler} field. \citet{Godoy2018} performed, for the first time, a search for wide binaries in the {\sl Kepler} field by using $RV$s and metallicities as criteria based on LAMOST-derived stellar parameters and the {\it Gaia} DR2. After examining data of five different surveys including the {\sl Kepler} mission, \citet{Moe2019} provided strong evidence with the help of LAMOST data that the close-binary fraction of solar-type stars is strongly anti-correlated with metallicity.  Those metallicities were taken from LAMOST data.

When constructing theoretical models for eclipsing binaries based on the light curves of {\sl Kepler} photometry, atmospheric parameters of the stars derived from the LK project have been used as key input parameters. While studying pulsating stars in individual binaries, \citet{Catanzaro2018} presented a spectroscopic and photometric analysis of the ellipsoidal variable star KIC~7599132 in a binary system in the {\sl Kepler} field. LAMOST spectra of KIC~5219533 triggered the discovery that this star is actually a hierarchical SB3 system. \citet{Catanzaro2019} derived the orbits and the atmospheric parameters of the inner SB2 pair, two twin Am stars, and evaluated some physical properties of the third component.
\citet{Zhang2018c, Zhang2019, Zhang2020c} carried out a seismic study of the $\gamma$ Doradus-type pulsations in the eclipsing binary KIC~10486425, reported the discovery and seismic analysis of an EL CVn-type binary with hybrid $\delta$ Sct-$\gamma$ Dor pulsations, and demonstrated KIC~9850387 as a short-period PMS eclipsing binary consisting of a hybrid $\gamma$ Dor-$\delta$ Sct primary component that has a nearly non-rotating core. \citet{Chen2020} discovered an Algol-type eclipsing binary, KIC~10736223, that has just undergone the rapid mass-transfer stage, with six $\delta$ Scuti-type pulsation modes detected. For the study of other types of binaries, \citet{Wang2019} reported discovery of two new R CMa-type eclipsing binaries containing a possible low-mass Helium white dwarf precursor: KIC~7368103 and KIC~8823397. \citet{Liu2020} presented a comprehensive photometric investigation of the active early K-type contact system IL Cancri.

\section{Phase~II of the Survey}
\label{sect:6}
From 2018, the LK-MRS project, a parallel project to LK, was approved to use the medium-resolution ($R\sim7500$) LAMOST spectrographs to observe objects in four central LK plates, during the bright nights in each lunar month. In contrast to the LK project that attempted to observe as many targets as possible with the wavelength range 370 to 900\,nm, the LK-MRS project aims to collect time-domain medium-resolution spectra for 20 plates (4 {\sl Kepler} and 16 {\sl K}2 plates) simultaneously in two spectral windows, 495-535\,nm and 630-680\,nm, over a period of 5 years. This strategy will provide high-quality spectra that will enable, for instance, the discovery of new binary stars and the study of their orbital properties \citep{2010ApJS..190....1R}, the characterization of the properties of high-amplitude (radial order) pulsating stars \citep{2008AcA....58..193S} and the monitoring of the variability of active stars \citep{Frasca2016}.

\begin{figure}
\centering
\includegraphics[width=10cm]{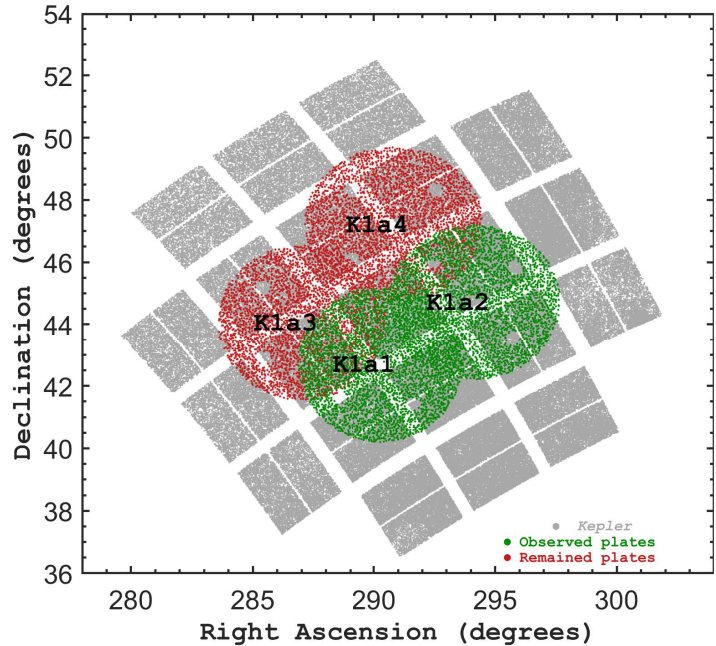}
\caption{Sky coverage of footprints from the LK-MRS project stamped over the targets in the {\sl Kepler} field.
\label{rd}}
\end{figure}

According to the time allocation, LK-MRS will obtain spectra for the four central LK plates with each plate observed about 60 times \citep[see details in][]{2020arXiv200507210L}. Their location is shown in Figure\,\ref{rd}, labeled as K1a1, K1a2, K1a3 and K1a4. Those plates contain about 12,000 stars down to $g\sim15.5$~mag, with $\sim75$\% of them observed by {\sl Kepler}. The actual observation sequence is determined by an automatic Python code which is used to decide which plate has the top priority. Once the plate is chosen, it will be continuously observed until it goes out of the LAMOST view \citep[see details in][]{zong2020}.

\begin{figure}
\centering
\includegraphics[width=10cm]{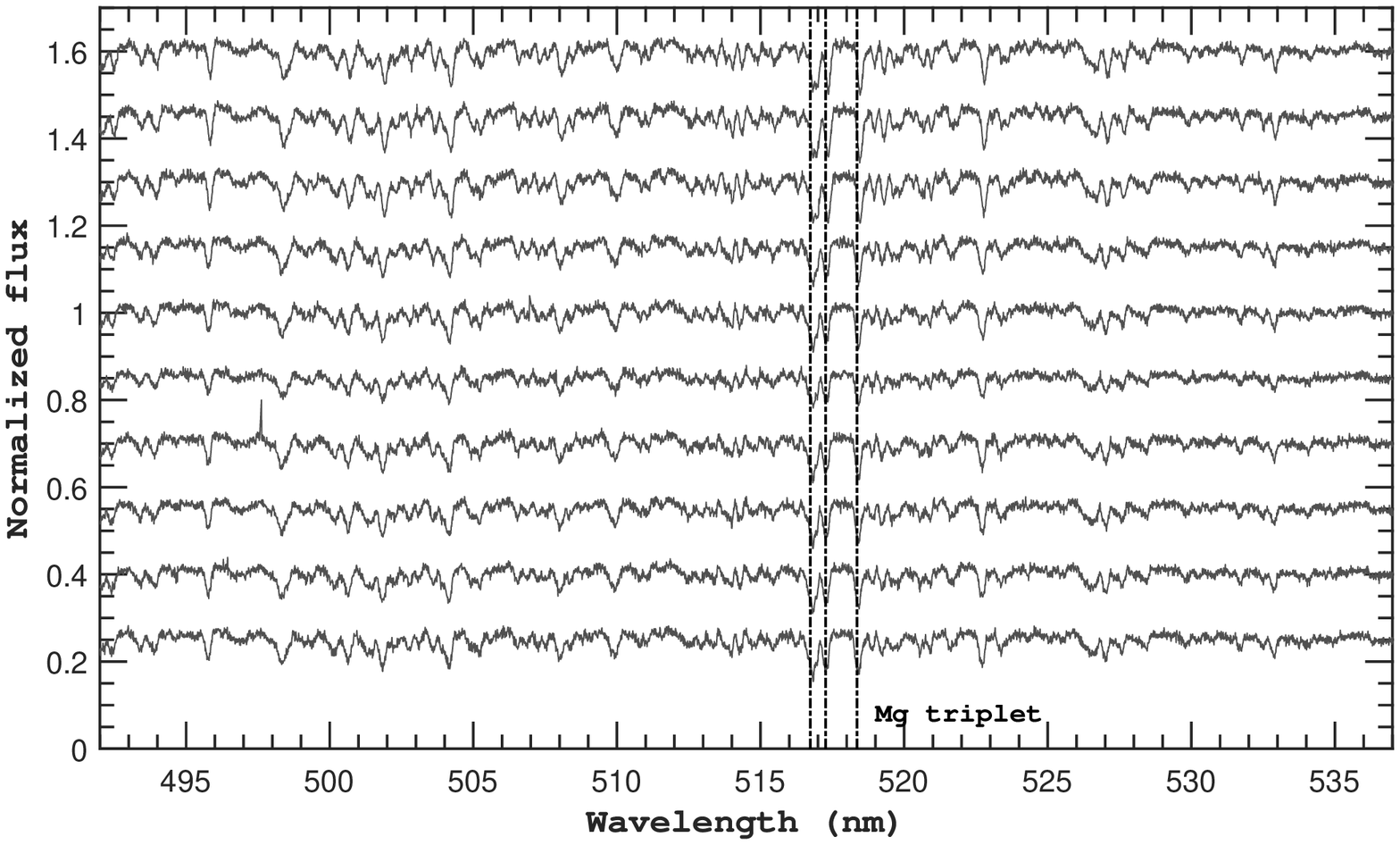}
\caption{Examples of LAMOST medium-resolution spectra of the object KIC\,08685306 in blue band. These ten time-domain spectra were obtained in two nights from the LK-MRS survey, with quality of $S/N \sim 50$. Flux of these spectra is normalized and shifted with relative values for visualization. The most distinct absorption lines for instance the Magnesium (Mg) triplet are marked with vertical lines.}
\label{spectra}
\end{figure}

Up to 2019 June, for two out of the four plates, namely K1a1 and K1a2, 39 and 4 visits were made, respectively.
The data processing of medium-resolution spectra is similar to that of the low-resolution ones, but the higher spectral resolution is taken into account. A total of 76,921 spectra for 4,578 objects have been collected for the K1a1 and K1a2 plates. Figure\,\ref{spectra} shows an example of the normalized spectra in the blue spectral range for an eclipsing binary star observed by {\sl Kepler}, KIC\,08685306, where the Mg triplet lines ($\lambda \sim 517$\,nm) are clearly seen.

The LASP pipeline provides the atmospheric parameters ($T_\mathrm{eff}$, $\log g$, and [Fe/H]), and radial velocity ($RV$) for type late-A, F, G and K stars through the analysis of the blue-arm medium-resolution spectra. Currently, a total of 71,914 groups of parameters of 3,981 stars are provided for the K1a1 and K1a2 plates. Other parameters such as $v \sin i$ and [$\alpha/$Fe] can also be provided but their quality is still under investigation. The $RV$s have small offsets that differ from spectrograph to spectrograph and also show a slight dependence on observation time, but these can be eliminated easily (Liu et al. 2019). As the objects are re-visited multiple times, those measurements can be used to estimate the internal uncertainties of LASP parameters and $RV$s. \citet{zong2020} finds that $T_\mathrm{eff}$, $\log g$, [Fe/H] and $RV$ have internal uncertainties of 100\,K, 0.15\,dex, 0.09\,dex and 1.00\,km/s respectively for S/N $= 10$. The precisions increase as S/N increases but they approach stable values for S/N $> 50$ of $\sim30$\,K, $\sim0.04$\,dex, $\sim0.02$\,dex and $\sim0.75$\,km/s, respectively.

Parameters from three other large surveys are used as external calibrators for LK-MRS by \citet{zong2020}. Their results suggest that the parameters derived from LK-MRS spectra generally agree well with those from the LK project and APOGEE, but the scatter increases as $\log g$ decreases when comparing with the APOGEE measurements. A large $T_\mathrm{eff}$ discrepancy is found with the {\it Gaia} values. The $RV$ comparisons of LK-MRS to {\it Gaia} and to APOGEE follow gaussian distributions with $\mu\sim1.10$ and $0.73$\,km\,s$^{-1}$, respectively.

These data are particularly important for discovering new binaries through $RV$ variations. A tentative simulation led by Wang et al. (in prep.) based on K1a1 observations suggests that the percentage of binary systems is higher than 10\%, roughly 200 out of 1900 stars, which is about ten times the fraction of  eclipsing binaries. To complement the high-precision {\sl Kepler} photometry, the time-series $RV$s are also useful in the determination of the physical properties of high-amplitude radial-order pulsators. Wang et al. (in prep.) is constructing seismic models for an RR Lyrae star with the $RV$s and {\sl Kepler} photometry.

\section{Summary and Prospects}
\label{sect:conclusions}
With the aim of collecting low-resolution (R$\approx$1800) spectra for as many stars as possible in the {\sl Kepler} field of view, phase~I of the LAMOST-{\sl Kepler} project began in 2010, while observations producing "qualified" spectra were started in 2012 June. Included in LAMOST DR7 (released March 2020), 238,386 low-resolution spectra with SNR$_g\geq6$ have been collected for 155,623 stars in the {\sl Kepler} field, in which 84,976 stars are in common with those observed by the {\sl Kepler} mission constituting $\sim43\%$ of the {\sl Kepler} targets. Stellar parameters have been derived and spectral classification carried out by three teams with the analysis codes LASP, ROTFIT and MKCLASS, respectively.

A wide use of LK and {\sl Kepler} data by the scientific community is witnessed by the large number of papers in the literature. By making a preliminary  search, we have found 70 refereed publications from 2014 to 2020 and a similar number of non-refereed articles. We divide the former into six research areas to summarize the research progress obtained with the help of the LK project data.

In the autumn of 2018, phase~II of the LK project, also called LK-MRS, started observing about 12,000 stars in four central LK-fields, aiming to complete about 60 visits of each target star in multiple epochs in five years. The currently available work shows that the LK-MRS observations have resulted in high-quality data which will prove to be useful for research in the fields of binary stars, high-amplitude pulsating stars, etc.

Both the LK and LK-MRS observations are continuing along with other LAMOST surveys. The pipelines are working well and provide large amounts of high-quality spectra, as well as atmospheric parameters, radial velocities, projected rotational velocities and chemical abundances of more than 10 elements for a great number of stars in the {\sl Kepler} field. We encourage astronomers to use the LK and LK-MRS data to carry out scientific research.  We expect an increasing number of papers using LK data will appear in the future. 

\begin{acknowledgements}
The Guoshoujing Telescope (the Large Sky Area Multiobject Fiber Spectroscopic Telescope LAMOST) is a National Major Scientific Project built by the Chinese Academy of Sciences. Funding for the project has been provided by the National Development and Reform Commission. LAMOST is operated and managed by the National Astronomical Observatories, Chinese Academy of Sciences. JNF and WKZ acknowledge the support from National Natural Science Foundation of China (NSFC) through the grant 11833002 and the Beijing Natural Science Foundation (No. 1194023)..
\end{acknowledgements}


\begin{thebibliography}{99}
\small \setlength{\itemindent}{-3mm}
\setlength{\itemsep}{-0.5mm}
\setlength{\baselineskip}{4.7mm}

\bibitem[{Balona} (2018)]{Balona2018} Balona, L. A.; 2018, MNRAS, 476, 4840

\bibitem[{Barentsen} {et~al.}(2018)]{Barentsen2018}
{Barentsen}, G.; {Hedges}, C.; {Saunders}, N.; {et~al.} 2018, arXiv:1810.12554


\bibitem[{Bashi \& Zucker} (2019)]{Bashi2019}
Bashi, D. \& Zucker, S. 2019, AJ, 158, 61
\bibitem[{Batalha} {et~al.}(2010)]{Batalha2010}
Batalha, N. M.; Borucki, W. J.; Koch, D. G.; et al. 2010, ApJL, 713, L109
\bibitem[{Bell} {et~al.}(2017)]{Bell2017}
Bell, K.J.; Gianninas, A.; Hermes, J.J.; et al. 2017, ApJ, 835, 180

\bibitem[{Bhardwaj} {et al.}(2019)]{Bhardwaj2019}
Bhardwaj, A.; Kanbur, S.; He, H.; et al. 2019, ApJ, 884, 20
\bibitem[{Borucki} {et~al.}(2010)]{Borucki2010}
{Borucki}, W.J.; {Koch}, D.; {Basri}, G.; {et~al.} 2010, Science, 327, 977

\bibitem[{Brown} {et~al.}(2011)]{Brown2011} Brown, T.M.; Latham, D.W.; Everett, M.E.; et al. 2011, AJ, 142, 112
\bibitem[{Casey} {et~al.}(2018)]{Casey2018} Casey, A. R.; Kennedy, G.M.; Hartle, T.R.; et al. 2018, MNRAS, 478, 2812
\bibitem[{Catanzaro} {et~al.}(2018)]{Catanzaro2018} Catanzaro, G.; Frasca, A.; Giarrusso, M.; et al. MNRAS, 2018, 477, 2020
\bibitem[Catanzaro et al.(2019)]{Catanzaro2019} Catanzaro, G., Gangi, M., Giarrusso, M., et al.\ 2019, \mnras, 487, 919
\bibitem[{Chang} {et~al.}(2017)]{Chang2017} Chang, H.Y.; Song, Y.H.; Luo, A.L.; et al. 2017, ApJ, 834, 92
\bibitem[{Chang} {et~al.}(2018)]{Chang2018} Chang, H.Y.; Lin, C.L.; Ip, W.H.; et al. 2018, ApJ, 867, 78
\bibitem[{Chen} {et~al.}(2020)]{Chen2020} Chen, X.H.; Zhang, X,B,; Li, Y.; et al. 2020, ApJ, 895, 136
\bibitem[{Cook} {et~al.}(2017)]{Cook2017} Cook, N.J.; Pinfield, D.J.; Marocco, F.; et al. 2017, MNRAS, 467, 5001
\bibitem[{Cui} {et~al.}(2019)]{Cui2019} Cui, K.M.; Liu, J.F.; Yang, S.H.; et al. 2019, MNRAS, 489, 5513
\bibitem[{Cui} {et~al.}(2012)]{Cui2012} Cui, X.Q.; Zhao, Y.H.; Chu, Y.Q.; et al. 2012, RAA, 12, 1197
\bibitem[{Cunha} {et~al.}(2007)]{Cunha2007} Cunha M.S., Aerts; C.; Christensen-Dalsgaard, J.; et al. 2007, A\&ARv, 14, 217
\bibitem[{De Cat} {et~al.}(2015)]{De Cat2015} De Cat, P.; Fu, J.N.; Ren, A.B.; et al. 2015, ApJS, 220, 19
\bibitem[{Dong} {et~al.}(2014)]{Dong2014} Dong, S.B.; Zheng, Z.; Zhu, Z.; et al. 2014, ApJL, 789, L3
\bibitem[{Dong} {et~al.}(2018)]{Dong2018} Dong, S.B.; Xie, J.W.; Zhou, J.L.; et al. 2018, PNAS, 115, 266
\bibitem[{Frasca} {et~al.}(2003)]{Frasca2003} Frasca, A.; Alcal¨¢, J.M.; Covino, E.; et al. 2003, A\&A, 405, 149
\bibitem[{Frasca} {et~al.}(2006)]{Frasca2006} Frasca, A.; Guillout, P.; Marilli, E.; et al. 2006, A\&A, 454, 301
\bibitem[Frasca et al.(2016)]{Frasca2016} Frasca, A., Molenda-{\.Z}akowicz, J., De Cat, P., et al.\ 2016, \aap, 594, A39
\bibitem[{Godoy-Rivera \& Chaname}(2018)]{Godoy2018} Godoy-Rivera, D. \& Chaname, J. 2018, MNRAS, 479, 4440
\bibitem[{Gray} {\&}{ Corbally}(2014)]{Gray2014} Gray, R.O.; \& Corbally, C.J. 2014, AJ, 147, 80
\bibitem[{Gray} {et~al.}(2016)]{Gray2016} Gray, R.O.; Corbally, C.J.; De Cat, P.; et al. 2016, AJ, 151, 13
\bibitem[{Guo} {et~al.}(2017)]{Guo2017} Guo, X.Y.; Johnson, J.A.; Mann, A.W.; et al. 2017, ApJ, 838, 25
\bibitem[{Hardegree-Ullman} {et~al.}(2019)]{Hardegree2019} Hardegree-Ullman, K.K.; Cushing, M.C.; Muirhead, P.S.; et al. 2019, AJ, 158, 75
\bibitem[{Hey} {et~al.}(2019)]{Hey2019} Hey, D.R.; Holdsworth, D.L.; Bedding, T.R.; et al. 2019, MNRAS, 488, 18
\bibitem[{Ho} {et~al.}(2017)]{Ho2017} Ho, A.Y.Q.; Rix, H.-W.; Ness, M.K.; et al. 2017, ApJ, 841, 40
\bibitem[{Hou} {et~al.}(2018)]{Hou2018} Hou, Y.H.; Tang, L.L.; Xu, M.M.; et al. 2018, Proceedings of SPIE, 10702, 107021I
\bibitem[{Karoff} {et~al.}(2016)]{Karoff2016} Karoff, C.; Knudsen, M.F.; De Cat, P.; et al. 2016, Nature Comm. 7, 11058
\bibitem[{Kawahara} {\&} {Masuda} (2019)]{Kawahara2019} Kawahara, H; \& Masuda, K. 2019, AJ, 157, 218
\bibitem[{Kumar} {et~al.}(2018)]{Kumar2018} Kumar, Y.B.; Singh, R.; Reddy, B.E.; et al. 2018, ApJL, 858, L22
\bibitem[{Li} {et~al.}(2017)]{Li2017} Li, Y.G.; Du, M.H.; Xie, B.H.; et al. 2017, RAA, 17, 44
\bibitem[{Li} {et~al.}(2018)]{Li2018} Li, Y.G.; Bedding, T.R.; Li, T.D.; et al. 2018, MNRAS, 476, 470
\bibitem[{Liu} {et~al.}(2015)]{Liu2015} Liu, C.; Fang, M.; Wu, Y.; et al. 2015, ApJ, 807, 4
\bibitem[Liu et al.(2019)]{2019RAA....19...75L} Liu, N., Fu, J.N., Zong, W., et al.\ 2019, RAA, 19, 75.
\bibitem[{Liu} {et~al.}(2020)]{Liu2020} Liu, N.P.; Sarotsakulchai, T.; Rattanasoon, S.; et al. 2020, PASJ, doi:10.1093/pasj/psaa062
\bibitem[Liu et al.(2020)]{2020arXiv200507210L} Liu, C.; Fu, J.N.; Shi, J.R.; et al.\ 2020, arXiv:2005.07210
\bibitem[{Lu} {et~al.}(2019)]{Lu2019} Lu, H.P.; Zhang, L.Y.; Shi, J.R.; et al. 2019, ApJS, 243, 28
\bibitem[{Luo} {et~al.}(2012)]{Luo2012} Luo, A.L.; Zhang, H.T.; Zhao, Y.H.; et al. 2012, RAA, 12, 1243
\bibitem[{Luo} {et~al.}(2015)]{Luo2015} Luo, A.L.; Zhao, Y.H.; Zhao, G.; et al. 2015, RAA, 15, 1095
\bibitem[{Mathur} {et~al.}(2017)]{Mathur2017} Mathur, S.; Huber, D.; Batalha, N.M.; et al. 2017, ApJS, 229, 30
\bibitem[{McNamara} {et~al.}(2012)]{McNamara2012} McNamara, B.J.; Jackiewicz, J.; McKeever, J. 2012, AJ, 143, 101
\bibitem[{Michel} (2006)]{Michel2006}{Michel}, E.; 2006, CoAst, 147, 40

\bibitem[{Mints} {\&}{ Hekker}(2017)]{Mints2017} Mints, A.; \& Hekker, S. 2017, A\&A, 604, 108
\bibitem[{Moe} {et~al.}(2019)]{Moe2019} Moe, M.; Kratter, K.M.; Badenes, C.; 2019, ApJ, 875, 61

\bibitem[{Molenda-{\.Z}akowicz} {et~al.}(2010)]{Molenda2010} Molenda-{\.Z}akowicz, J.; Jerzykiewicz, M.; Frasca, A.; et al. 2010, arXiv:1005.0985

\bibitem[{Mulders} {et~al.}(2016)]{Mulders2016} Mulders, G.D.; Pascucci, I.; Apai, D.; et al. 2016, AJ, 152, 187
\bibitem[{Murphy} {et~al.}(2016)]{Murphy2016} Murphy, S.J.; Bedding, T.R.; Shibahashi, H. 2016, ApJL, 827, L17
\bibitem[{Narang} {et~al.}(2018)]{Narang2018} Narang, M.; Manoj, P.; Furlan, E.; et al. 2018, AJ, 156, 221
\bibitem[{Pande} {et~al.}(2018)]{Pande2018} Pande, D.; Bedding, T.R.; Huber, D.; et al. 2018, MNRAS, 480, 467
\bibitem[{Peralta} {et~al.}(2018)]{Peralta2018} Peralta, R.A.; Samadi, R.; Michel, E. 2018, AN, 339, 134
\bibitem[{Petigura} {et~al.}(2017)]{Petigura2017} Petigura, E.A.; Howard, A.W.; Marcy, G.W.; et al. 2017, AJ, 154, 107
\bibitem[{Petigura} {et~al.}(2018)]{Petigura2018} Petigura, E.A.; Marcy, G.W.; Winn, J.N.; et al. 2018, AJ, 155, 89
\bibitem[Raghavan et al.(2010)]{2010ApJS..190....1R} Raghavan, D., McAlister, H.~A., Henry, T.~J., et al.\ 2010, \apjs, 190, 1
\bibitem[{Ren} {et~al.}(2016a)]{Ren2016a} Ren, A. B.; Fu, J.N.; De Cat, P.; et al. 2016a, ApJS, 225, 28
\bibitem[{Ren} {et~al.}(2016b)]{Ren2016b} Ren, J.J.; Liu, X.W.; Xiang, M.S.; et al. 2016b, RAA, 16, 45
\bibitem[{Scaringi} {et~al.}(2018)]{Scaringi2018} Scaringi, S.; Knigge, C.; Drew, J.E.; et al. 2018, MNRAS, 481, 3357
\bibitem[{Silva Aguirre} {et~al.}(2014)]{Silva2014} Silva Aguirre, V.; Ruchti, G.R.; Hekker, S.; et al. 2014, ApJL, 784, L16
\bibitem[{Singh} {et~al.}(2019a)]{Singh2019a} Singh, R.; Reddy, B.E.; Kumar, Y.B.; et al. 2019a, ApJL, 878, L21
\bibitem[{Singh} {et~al.}(2019b)]{Singh2019b} Singh, R.; Reddy, B.E.; Kumar, Y.B.; 2019b, MNRAS, 482, 3822
\bibitem[{Smalley} {et~al.}(2017)]{Smalley2017} Smalley, B.; Antoci, V.; Holdsworth, D. L.; et al. 2017, MNRAS, 465, 2662
\bibitem[Smolec \& Moskalik(2008)]{2008AcA....58..193S} Smolec, R.; \& Moskalik, P.\ 2008, \actaa, 58, 193
\bibitem[{Stassun} {et~al.}(2018)]{Stassun2018} Stassun, K.G.; Corsaro, E.; Pepper, J.A.; et al. 2018, AJ, 1,22
\bibitem[{Wang} {et~al.}(2019)]{Wang2019} Wang, K.; Zhang, X.B.; Luo, Y.P.; et al. 2019, MNRAS, 486, 2462
\bibitem[{Wang} {et~al.}(2016)]{Wang2016} Wang, L.; Wang, W.; Wu, Y.; et al. 2016, AJ, 152, 6
\bibitem[{Wang} {et~al.}(1996)]{Wang1996} Wang, S.G.; Su, D.Q.; Chu, Y.-Q.; et al. 1996, ApOpt, 35, 5155
\bibitem[{Wang} {et~al.}(2018)]{Wang2018} Wang, W.; Wang, L.; Li, X.; et al. 2018, ApJ, 860, 136
\bibitem[{Wu} {et~al.}(2017)]{Wu2017} Wu, Y.Q.; Xiang, M.S.; Zhang, X.F.; et al. 2017, RAA, 17, 5
\bibitem[{Wu} {et~al.}(2018)]{Wu2018} Wu, Y.Q.; Xiang, M.S.; Bi, S.L.; et al. 2018, MNRAS, 475, 3633
\bibitem[{Wu} {et~al.}(2019)]{Wu2019} Wu, Y., Xiang, M., Zhao, G., et al.\ 2019, \mnras, 484, 5315
\bibitem[{Xiang} {et~al.}(2017)]{Xiang2017} Xiang, M.S.; Liu, X.W.; Shi, J.R.; et al. 2017, MNRAS, 464, 3657
\bibitem[{Xie} {et~al.}(2016)]{Xie2016} Xie, J.W.; Dong, S.B.; Zhu, Z.H.; et al. 2016, PNAS, 113, 11431
\bibitem[{Xing} {et~al.}(1998)]{Xing1998} Xing, X.; Zhai, C.; Du, H.; et al. 1998, Proc. SPIE, 3352, 839
\bibitem[{Yang \& Liu} (2019)]{Yang2019} Yang, H.Q.; \& Liu, J.F.; 2019, ApJS, 241, 29
\bibitem[{Yang} {et~al.}(2017)]{Yang2017} Yang, H.Q.; Liu, J.F.; Gao, Q.; et al., ApJ, 2017, 849, 36
\bibitem[{Yu} {et~al.}(2016)]{Yu2016} Yu, J.; Huber, D.; Bedding, T.R.; et al. 2016, MNRAS, 463, 1297
\bibitem[{Yu} {et~al.}(2018)]{Yu2018} Yu, J.; Huber, D.; Bedding, T.R.; et al. 2018, ApJS, 236, 42
\bibitem[{Zhang} {et~al.}(2018a)]{Zhang2018a} Zhang, C.G.; Liu, C.; Wu, Y.; et al. 2018a, ApJ, 854, 168
\bibitem[{Zhang} {et~al.}(202a0)]{Zhang2020a} Zhang, J.H.; Shapiro, A.; Bi, S.L.; et al. 2020a, ApJL, 894, L11
\bibitem[Zhang et al.(2020b)]{Zhang2020b} Zhang, J.H., Bi, S.L., Li, Y.G., et al.\ 2020b, \apjs, 247, 9
\bibitem[{Zhang} {et~al.}(2018b)]{Zhang2018b} Zhang, L.Y.; Lu, H.P.; Han, X.M.; et al. 2018b, New Astron., 61, 36
\bibitem[{Zhang} {et~al.}(2018c)]{Zhang2018c} Zhang, X.B.; Fu, J.N.; Luo, C.Q.; et al. 2018c, ApJ, 865, 115
\bibitem[{Zhang} {et~al.}(2019)]{Zhang2019} Zhang, X.B.; Wang, K.; Chen, X.H.; et al. 2019, ApJ, 884, 165
\bibitem[{Zhang} {et~al.}(2020c)]{Zhang2020c} Zhang, X.B.; Chen, X.H.; Zhang, H.T.; et al. 2020c, ApJ, 895, 124
\bibitem[{Zhu} {et~al.}(2018)]{Zhu2018} Zhu, W.; Petrovich, Cr.; Wu, Y.Q; et al. 2018, ApJ, 860, 101
\bibitem[{Zhu}(2019)]{Zhu2019} Zhu, W. 2019, ApJ, 873, 8
\bibitem[{Zong} {et~al.}(2018)]{Zong2018} Zong, W.; Fu, J.N.; De Cat, P.; et al. 2018, ApJS, 238, 30
\bibitem[Zong et al.(2020)]{zong2020} Zong, W.; Fu, J.N.; De Cat, P.; et al.\ 2020, \apjs, submitted


%

\end{thebibliography}

\end{document}